\documentclass[preprint,showpacs,preprintnumbers,amsmath,amssymb]{revtex4}

\usepackage{graphicx}
\usepackage{dcolumn}
\usepackage{bm}

\begin{document}

\title{Schemes of implementation in NMR of quantum processors and Deutsch-Jozsa algorithm by using virtual spin representation}
\author{Alexander R. Kessel}
\email{kessel@dionis.kfti.knc.ru}
\author{Natalia M. Yakovleva}
\affiliation{Kazan Physical-Technical Institute, Russian Academy of Science, Sibirsky trakt 10/7, Kazan 420029, Russia}
\date{\today}

\begin{abstract}
Schemes of experimental realization of the main two qubit processors for quantum computers and Deutsch-Jozsa algorithm   are derived in virtual spin representation. The results are applicable for every four quantum states allowing the required properties for quantum processor implementation if for qubit encoding virtual spin representation is used. Four dimensional Hilbert space of nuclear spin $3/2$ is considered in details for this aim.
\end{abstract}

\maketitle

\section{Introduction}
Liquids, containing molecules with a few interacting spins $1/2$, are evident leaders in experimental implementation of quantum processors and Deutsch-Jozsa \cite{deutsch1985}, Grover \cite{grover1997} and Shor \cite{shor1997} algorithms up to day. In spite of this fact an analyses of quantum information achievements, which was performed by several investigation groups \cite{jones2000,divincenzo2000,cory2000,valiev2001} shows that possibilities limit of such systems will be reached in the nearest future. In a big review article \cite{cory2000} sixteen authors well known in quantum informatics came to a conclusion that the next generation of quantum processors will be built on a quadrupole nuclei with spins $I>1/2$ in solid media.

Some time earlier the problem of qubit coding in arbitrary system of quantum states was solved by means of introduction of virtual spin representation \cite{kessel1999a}. For the use of these states in quantum informatics one has to be able to excite resonance transitions for experimental implementation of quantum processors. In particular these states may be that of mentioned in Ref.~\cite{cory2000}.

The point of approach suggested in Ref.~\cite{kessel1999a} can be clarified by considering four states of nuclear spin $I=3/2$. Two real spin $R=1/2$ and $S=1/2$ for construction of two qubit logic elements in the standard model of quantum computer \cite{jones2000,divincenzo2000,cory2000,valiev2001} are used usually. In quantum mechanical formalism the states of such a system and transition between these states are usually described in an abstract four-dimensional space, which is a direct product  $\Gamma_{R}\otimes\Gamma_{S}$ of two-dimensional spaces of real spins $R$ and $S$ spaces. An inverse procedure was proposed in Ref. \cite{kessel1999a}: four-dimensional space $\Gamma_{I}$ of real spin $I=3/2$ was presented as a direct product $\Gamma_{R}\otimes\Gamma_{S}$ of two abstract two-dimensional state spaces of virtual spins equal to $1/2$. So every operator $\mathbf{P}$ being determined in four-dimensional basis $\Gamma_{I}$ can be expressed as a linear combination of products $\mathbf{R}\otimes\mathbf{S}$ of spin vector component operators, determined in $\Gamma_{R}$ and $\Gamma_{S}$ spaces. In other words, it is proposed to experimentally influence on the real spin $3/2$ for quantum processor implementation, but the logical sense of these actions to read from transformations of the virtual spin states.

Qubit coding in virtual spin representation has some preferences. This representation gives higher information recording density and calculation basis stability \cite{kess3}. One needs not stationary influence a spin system by a complicated sequences of radio frequency pulses for damping the spin exchange interaction and also needs not make windows for this interaction acting in the exactly given time interval. Gate independence of the spin exchange interaction brings it to a full experimenter control, and as a result to higher velocity of gate operation. (Independence of exchange interaction gives additional possibility for choosing substances for gate realization in virtual spin representation). Spin $3/2$ quadrupole interaction in solids may produce resonance frequency differences which are much greater than those due to exchange interaction in liquids. This fact makes it easier to address radio frequency pulses to individual resonance transition that is to individual virtual qubit. More shorter relaxation times and more rare spreading of quadrupole nuclei in comparison with $^{1}$H and $^{13}$C nuclei can be considered as difficulties of quadrupole nuclei use in quantum informatics.

Realization schemes of two qubit gates and Deutsch-Jozsa \cite{deutsch1985} algorithms in the nuclear spin $3/2$ states space $\Gamma_{I}$ where two virtual qubits are embedded will be presented below. The most suitable mathematical formalism for considering spin states corresponding to not equally spaced energy levels is the projection operator formalism. In case of spin $I$ projective operator $\mathbf{I}_{mn}$ is a $(2I+1)\times(2I+1)$ matrix with every  element $\mathbf{I}_{kl}$ equals to zero except for $\mathbf{I}_{mn}=1$.  Projective operators possess extremely simple multiplying rules
\begin{equation}
 \mathbf{I}_{kl}\mathbf{I}_{mn}=\delta_{lm}\mathbf{I}_{kn}, \label{equ1_1}
\end{equation}
as well as action rules on the basis functions of the Hilbert
space under consideration
\begin{equation}
 \mathbf{I}_{mn}|\Psi_{k}\rangle=\delta_{nk}|\Psi_{m}\rangle. \label{equ1_2}
\end{equation}
Spin operator components are expressed through projective operators in the following manner
\begin{equation}
 \mathbf{I}_{\alpha}=\sum_{m,n}\langle\Psi_{m}|\mathbf{I}_{\alpha}|\Psi_{n}\rangle \mathbf{I}_{mn}. \label{equ1_3}
\end{equation}
The formulae (\ref{equ1_2}) allows to consider $\mathbf{I}_{kn}$ as projective operators: arbitrary state $\sum C_{m}|\Psi_{m}\rangle$ is transformed in $C_{n}|\Psi_{k}\rangle$ or into ort $|\Psi_{k}\rangle$ of the Hilbert space under consideration under the action of operator $\mathbf{I}_{kn}$.

\begin{figure}
\includegraphics{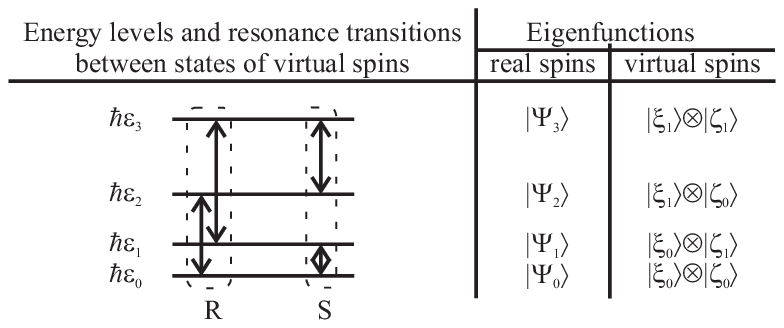}
\caption{\label{fig:f1}Energy levels, eigenstates of real spin $I=1/2$ and corresponding direct products of virtual spin states The transitions between states of virtual spin $R(S)$ in condition when virtual spin $S(R)$ is remained invariable is shown by arrows in dashed rectangles}
\end{figure}

A quadrupole nucleus with spin $3/2$, placed in dc magnetic field and crystalline field of low symmetry, has four not equally spaced energy levels $\hbar\varepsilon_{0}<\hbar\varepsilon_{1}<\hbar\varepsilon_{2}<\hbar\varepsilon_{3}$ (Fig.~\ref{fig:f1}). These states in the energetic representation are described by the Hamiltonian
\begin{equation}
 \mathcal{H}=\hbar\sum_{m} \varepsilon_{m}\mathbf{I}_{mm}. \label{equ1_4}
\end{equation}

For simplicity indices $0, 1, 2, 3$ will be written instead of spin $z$-component eigenvalues $m=-3/2, -1/2, +1/2, +3/2$, respectively. The functions $|\Psi_{\alpha}\rangle$ in Fig.~\ref{fig:f1} transform into $|\chi_{m}\rangle$ according to law
\begin{eqnarray}
& |\Psi_{0}\rangle \Rightarrow |\chi_{-3/2}\rangle, & |\Psi_{1}\rangle \Rightarrow |\chi_{-1/2}\rangle, \nonumber \\
& |\Psi_{2}\rangle \Rightarrow |\chi_{1/2}\rangle, &  |\Psi_{3}\rangle \Rightarrow |\chi_{3/2}\rangle, \label{equ1_5}
\end{eqnarray}
if quadrupole interaction is neglected. Here $|\chi_{m}\rangle$ is eigenfunction, corresponding to eigenvalue $m$ of $\mathbf{I}_{z}$-operator.

Every operator given in a four-dimensional basis can be expressed as a linear combination of a direct product $\mathbf{R}\otimes \mathbf{S}$ of virtual spin vector components, given in subspaces $\Gamma_{R}$ and $\Gamma_{S}$. The following isomorphic correspondence between the basis $|\Psi_{M}\rangle$ of the space $\Gamma_{I}$ and the basis $|\xi_{m}\rangle \otimes |\zeta_{n}\rangle$ of the direct product of  the virtual spin spaces takes place
\begin{eqnarray}
 |\Psi_{0}\rangle = |\xi_{0}\rangle \otimes |\zeta_{0}\rangle \equiv |00 \rangle, & |\Psi_{1}\rangle = |\xi_{0}\rangle \otimes |\zeta_{1}\rangle \equiv |01 \rangle, \nonumber \\
 |\Psi_{2}\rangle = |\xi_{1}\rangle \otimes |\zeta_{0}\rangle \equiv |10 \rangle, &  |\Psi_{3}\rangle = |\xi_{1}\rangle \otimes |\zeta_{1}\rangle \equiv |11 \rangle, \label{equ1_6}
\end{eqnarray}
where the indices $1$ and $0$ are used for values $+1/2$ and $-1/2$ of virtual spins $z$-components. Here $|11 \rangle, |10 \rangle, \ldots$ are notations, which are usually used in the information theory for presenting two qubit states.

\section{Expressions for gates in $\Gamma_{I}$ and  $\Gamma_{R} \otimes \Gamma_{S}$ spaces}
The expressions for the most frequent one and two qubit gates in four-dimensional space $\Gamma_{I}$ of the real spin $3/2$ states and also in a direct product $\Gamma_{R} \otimes \Gamma_{S}$ of virtual spins $R$ and $S$ spaces will be given below. \\
Unity transformation
\begin{equation}
\bm{\mathcal{P}}_{1} \equiv \mathbf{E} = |\xi , \zeta \rangle \Rightarrow  |\xi , \zeta \rangle = \mathbf{I}_{00} + \mathbf{I}_{11} + \mathbf{I}_{22} + \mathbf{I}_{33} = \mathbf{e}_{R} \otimes \mathbf{e}_{S}, \label{equ2_7}
\end{equation}
where $\mathbf{E}$ is the unity operator in $\Gamma_{I}$, $\mathbf{e}_{R}$ and $\mathbf{e}_{S}$ are that of in $\Gamma_{R}$ and $\Gamma_{S}$. \\
Negation transformation in $\Gamma_{R}$
\begin{equation}
\bm{\mathcal{P}}_{2} \equiv \mathbf{NOT}_{1} = |\xi,\zeta\rangle \Rightarrow  |\xi',\zeta\rangle = \mathbf{I}_{02} + \mathbf{I}_{13} + \mathbf{I}_{20} + \mathbf{I}_{31} = \mathbf{r}_{x} \otimes \mathbf{e}_{S}, \label{equ2_8}
\end{equation}
where $\xi' = \neg \xi$ and $\mathbf{r}_{x} = 2\mathbf{R}_{x} \equiv \mathbf{r}_{01} + \mathbf{r}_{10}$ is the $x$-component of Pauli operator in $\Gamma_{R}$. \\
Negation transformation in $\Gamma_{S}$
\begin{equation}
\bm{\mathcal{P}}_{3} \equiv \mathbf{NOT}_{2} = |\xi , \zeta \rangle \Rightarrow  |\xi , \zeta' \rangle = \mathbf{I}_{01} + \mathbf{I}_{10} + \mathbf{I}_{23} + \mathbf{I}_{32} = \mathbf{e}_{R} \otimes \mathbf{s}_{x}, \label{equ2_9}
\end{equation}
where $\mathbf{s}_{x} = 2\mathbf{S}_{x} \equiv \mathbf{s}_{01} + \mathbf{s}_{10}$  is the $x$-component of Pauli operator in $\Gamma_{S}$. \\
Negation transformation in both $\Gamma_{R}$ and $\Gamma_{S}$ spaces
\begin{equation}
\bm{\mathcal{P}}_{4} \equiv \mathbf{NOT} = |\xi,\zeta\rangle \Rightarrow |\xi',\zeta'\rangle = \mathbf{I}_{03} + \mathbf{I}_{12} + \mathbf{I}_{21} + \mathbf{I}_{30} = \mathbf{r}_{x} \otimes \mathbf{s}_{x}. \label{equ2_10}
\end{equation}
Virtual spin states exchange
\begin{equation}
\bm{\mathcal{P}}_{5}\equiv\mathbf{SWAP}=|\xi,\zeta\rangle \Rightarrow |\zeta,\xi\rangle = \mathbf{I}_{00} + \mathbf{I}_{12} + \mathbf{I}_{21} + \mathbf{I}_{33} = (1/2)\mathbf{e}_{R} \otimes \mathbf{e}_{S} + (1/2)[\mathbf{r}_{x} \otimes \mathbf{s}_{x} + \mathbf{r}_{y} \otimes \mathbf{s}_{y} + \mathbf{r}_{z} \otimes \mathbf{s}_{z}]. \label{equ2_11}
\end{equation}
Controlled negation in $\Gamma_{S}$ (negation in $\Gamma_{S}$ space when virtual spin $R$ is in the state $|1 \rangle$)
\begin{equation}
\bm{\mathcal{P}}_{6} \equiv \mathbf{CNOT}_{1\rightarrow 2} = |\xi , \zeta \rangle \Rightarrow  |\xi, \xi \oplus \zeta \rangle = \mathbf{I}_{00} + \mathbf{I}_{11} + \mathbf{I}_{23} + \mathbf{I}_{32} = \mathbf{r}_{00} \otimes \mathbf{e}_{S} + \mathbf{r}_{11} \otimes \mathbf{s}_{x}, \label{equ2_12}
\end{equation}
where $\mathbf{r}_{mn}$ is a projective operator in $\Gamma_{R}$. \\
Controlled negation in $\Gamma_{R}$ (negation in $\Gamma_{R}$ space, when virtual spin $S$ is in the state $|1 \rangle$)
\begin{equation}
\bm{\mathcal{P}}_{7} \equiv \mathbf{CNOT}_{2\rightarrow 1} = |\xi , \zeta \rangle \Rightarrow  |\xi \oplus \zeta , \zeta \rangle = \mathbf{I}_{00} + \mathbf{I}_{13} + \mathbf{I}_{22} + \mathbf{I}_{31} = \mathbf{e}_{R} \otimes \mathbf{s}_{00} + \mathbf{r}_{x} \otimes \mathbf{s}_{11}, \label{equ2_13}
\end{equation}
where $\mathbf{s}_{mn}$ is a projective operator in $\Gamma_{S}$. \\
Inverse controlled negation in $\Gamma_{S}$ (negation in $\Gamma_{S}$ space, when virtual spin $R$ is in the state $|0 \rangle$)
\begin{equation}
\bm{\mathcal{P}}_{8} \equiv \mathbf{ICNOT}_{1\rightarrow 2} = |\xi , \zeta \rangle \Rightarrow  |\xi, \xi' \oplus \zeta' \rangle = \mathbf{I}_{01} + \mathbf{I}_{10} + \mathbf{I}_{22} + \mathbf{I}_{33} = \mathbf{r}_{00} \otimes \mathbf{s}_{x} + \mathbf{r}_{11} \otimes \mathbf{e}_{S}. \label{equ2_14}
\end{equation}
Inverse controlled negation in $\Gamma_{R}$ (negation in $\Gamma_{R}$ space, when virtual spin $S$ is in the state $|0 \rangle$)
\begin{equation}
\bm{\mathcal{P}}_{9} \equiv \mathbf{ICNOT}_{2\rightarrow 1} = |\xi , \zeta \rangle \Rightarrow  |\xi' \oplus \zeta', \zeta \rangle = \mathbf{I}_{02} + \mathbf{I}_{11} + \mathbf{I}_{20} + \mathbf{I}_{33} = \mathbf{r}_{x} \otimes \mathbf{s}_{00} + \mathbf{e}_{R} \otimes \mathbf{s}_{11}. \label{equ2_15}
\end{equation}
One qubit Hadamard operator are
\begin{eqnarray}
_{1}\mathbf{H}_{R} = (1/ \sqrt{2})[\mathbf{r}_{00} + \mathbf{r}_{01} + \mathbf{r}_{10} - \mathbf{r}_{11}] \otimes \mathbf{e}_{S}, \nonumber \\
_{1}\mathbf{H}_{S} = \mathbf{e}_{R} \otimes (1/ \sqrt{2})[\mathbf{s}_{00} + \mathbf{s}_{01} + \mathbf{s}_{10} - \mathbf{s}_{11}] \label{equ2_16}
\end{eqnarray}
or in $\Gamma_{I}$
\begin{eqnarray*}
_{1}\mathbf{H}_{R} = (1/ \sqrt{2})[\mathbf{I}_{00} + \mathbf{I}_{01} + \mathbf{I}_{10} - \mathbf{I}_{11} +  \mathbf{I}_{22} + \mathbf{I}_{23} + \mathbf{I}_{32} - \mathbf{I}_{33}], \\
_{1}\mathbf{H}_{S} = (1/ \sqrt{2})[\mathbf{I}_{00} + \mathbf{I}_{02} + \mathbf{I}_{11} + \mathbf{I}_{13} +  \mathbf{I}_{20} - \mathbf{I}_{22} + \mathbf{I}_{31} - \mathbf{I}_{33}].
\end{eqnarray*}
In many cases one can use pseudo Hadamard operator
\begin{eqnarray}
_{1}\mathbf{h}_{R} = (1/ \sqrt{2})[\mathbf{r}_{00} - \mathbf{r}_{01} + \mathbf{r}_{10} + \mathbf{r}_{11}] \otimes \mathbf{e}_{S}, \nonumber \\
_{1}\mathbf{h}_{S} = \mathbf{e}_{R} \otimes (1/ \sqrt{2})[\mathbf{s}_{00} - \mathbf{s}_{01} + \mathbf{s}_{10} + \mathbf{s}_{11}], \label{equ2_17}
\end{eqnarray}
instead of Hadamard operator $_{1}\mathbf{H}$ or in $\Gamma_{I}$
\begin{eqnarray*}
_{1}\mathbf{h}_{R} = (1/ \sqrt{2})[\mathbf{I}_{00} - \mathbf{I}_{01} + \mathbf{I}_{10} + \mathbf{I}_{11} +  \mathbf{I}_{22} - \mathbf{I}_{23} + \mathbf{I}_{32} + \mathbf{I}_{33}], \\
_{1}\mathbf{h}_{S} = (1/ \sqrt{2})[\mathbf{I}_{00} - \mathbf{I}_{02} + \mathbf{I}_{11} - \mathbf{I}_{13} +  \mathbf{I}_{20} + \mathbf{I}_{22} + \mathbf{I}_{31} + \mathbf{I}_{33}].
\end{eqnarray*}
Two qubit Hadamard operator is
\begin{eqnarray}
_{2}\mathbf{H} = ({_{1}\mathbf{H}}_{R}) ({_{1}\mathbf{H}}_{S}) = (1/2)[\mathbf{I}_{00} + \mathbf{I}_{01} + \mathbf{I}_{02} + \mathbf{I}_{03} + \mathbf{I}_{10} - \mathbf{I}_{11} + \mathbf{I}_{12} - \mathbf{I}_{13} \nonumber \\
+ \mathbf{I}_{20} + \mathbf{I}_{21} - \mathbf{I}_{22} - \mathbf{I}_{23} + \mathbf{I}_{30} - \mathbf{I}_{31} - \mathbf{I}_{32} + \mathbf{I}_{33}] \label{equ2_18}
\end{eqnarray}
and two qubit pseudo Hadamard operator is
\begin{eqnarray}
_{2}\mathbf{h} = ({_{1}\mathbf{h}}_{R})({_{1}\mathbf{h}}_{S}) = (1/2)[\mathbf{I}_{00} - \mathbf{I}_{01} - \mathbf{I}_{02} + \mathbf{I}_{03} + \mathbf{I}_{10} + \mathbf{I}_{11} - \mathbf{I}_{12} - \mathbf{I}_{13} \nonumber \\
+ \mathbf{I}_{20} - \mathbf{I}_{21} + \mathbf{I}_{22} - \mathbf{I}_{23}
+ \mathbf{I}_{30} + \mathbf{I}_{31} + \mathbf{I}_{32} + \mathbf{I}_{33}]. \label{equ2_19}
\end{eqnarray}
A sign changing operator of state $|\Psi_{m} \rangle$ is
\begin{equation}
  \mathbf{\Pi}_{m} = \mathbf{E} -2\mathbf{I}_{mm}. \label{equ2_20}
\end{equation}
The Deutsch-Jozsa  problem operators in $\Gamma_{I}$ are
\begin{eqnarray}
 \mathbf{D}_{00} & = & \mathbf{I}_{00} + \mathbf{I}_{11} + \mathbf{I}_{22} + \mathbf{I}_{33} = \mathbf{e}_{R} \otimes \mathbf{e}_{S}, \nonumber \\
 \mathbf{D}_{01} & = & \mathbf{I}_{00} + \mathbf{I}_{11} + \mathbf{I}_{23} + \mathbf{I}_{32} = \mathbf{CNOT}_{1 \rightarrow 2}, \nonumber \\
 \mathbf{D}_{10} & = & \mathbf{I}_{10} + \mathbf{I}_{01} + \mathbf{I}_{22} + \mathbf{I}_{33} = \mathbf{ICNOT}_{1 \rightarrow 2}, \nonumber \\
 \mathbf{D}_{11} & = & \mathbf{I}_{01} + \mathbf{I}_{10} + \mathbf{I}_{23} + \mathbf{I}_{32} = \mathbf{NOT}_{2}. \label{equ2_21}
\end{eqnarray}

\section{\label{sec:experimental} Schemes of experimental realization of quantum processors}
Let a pulsed external alternating magnetic field
\begin{equation}
\mathbf{H}_{t}=y_{0}H_{1}\cos(\Omega t-f), \label{equ3_22}
\end{equation}
be applied to a NMR sample, where $y_{0}$ is a polarization ort and $H_{1}$ is a pulse field amplitude.
The Hamiltonian of the spin interaction with this field is
\begin{equation}
\mathcal{H}_{t}=-h \gamma H_{1}\cos(\Omega t-f)\mathbf{I}_{y}, \label{equ3_23}
\end{equation}
where $\gamma$ is a gyromagnetic  ratio. In condition of selective excitation at the resonance $(\Omega = \Omega_{mn})$ transition between any pair of energy levels $\hbar \varepsilon_{m} \leftrightarrow \hbar \varepsilon_{n}$ of an arbitrary physical system field (\ref{equ3_22}) leads to the following expression  for the evolution operator (propogator)
\begin{eqnarray}
\mathbf{U}_{mn}(\varphi,f)=\mathbf{E}-(\mathbf{I}_{nn} +
\mathbf{I}_{mm})2\sin^{2}(\varphi/4) +
(\mathbf{I}_{nm}e^{if}-\mathbf{I}_{mn}e^{-if})\sin(\varphi/2),
\label{equ3_24}
\end{eqnarray}
where $\varphi = \gamma H_{1}t_{i} | \langle \Psi_{m}|I_{x}|\Psi_{n} \rangle |$ and $t_{i}$ is a pulse duration. For four-level system evolution operator (\ref{equ3_24}) can be presented in another form
\begin{eqnarray}
\mathbf{Y}_{mn}(\varphi,f)= \mathbf{I}_{kk} + \mathbf{I}_{ll}
+(\mathbf{I}_{nn} + \mathbf{I}_{mm})\cos(\varphi/2) +
(\mathbf{I}_{nm}e^{if}-\mathbf{I}_{mn}e^{-if})\sin(\varphi/2),
\label{equ3_25}
\end{eqnarray}
where indices $k,l \neq m,n$ and
\begin{eqnarray}
\mathbf{X}_{mn}(\varphi,f) = \mathbf{I}_{kk} + \mathbf{I}_{ll} +(\mathbf{I}_{nn} + \mathbf{I}_{mm})\cos(\varphi/2) - i (\mathbf{I}_{mn}e^{if} + \mathbf{I}_{nm}e^{-if})\sin(\varphi/2), \label{equ3_26}
\end{eqnarray}
if $\mathbf{H}_{t}$ field is polarized along the $x$-axis or when
the phase in Eq. (\ref{equ3_25}) is shifted $f \rightarrow f +
\pi/2$.

A few useful special expressions for spin 3/2:
\begin{eqnarray}
\mathbf{X}_{mn}(\pi/2)= \mathbf{I}_{kk} + \mathbf{I}_{ll} + (1/\sqrt{2})(\mathbf{I}_{nn} + \mathbf{I}_{mm} - i (\mathbf{I}_{mn}+\mathbf{I}_{nm})), \label{equ3_27}
\end{eqnarray}
\begin{equation}
\mathbf{X}_{mn}(\pi)= \mathbf{I}_{kk} + \mathbf{I}_{ll} -i (\mathbf{I}_{mn}+\mathbf{I}_{nm}), \label{equ3_28}
\end{equation}
\begin{eqnarray}
\mathbf{X}_{mn}(\pi)\mathbf{X}_{kl}(\pi)= -i (\mathbf{I}_{kl}+\mathbf{I}_{lk}+ \mathbf{I}_{mn} + \mathbf{I}_{nm}), k,l\neq m,n. \label{equ3_29}
\end{eqnarray}
RF field (\ref{equ3_22}) (and also Hamiltonian (\ref{equ3_23}) and propagators (\ref{equ3_24}-\ref{equ3_26}) connected with it) are the arsenal (not full!), which an experimentalist has for an implementation of logic gates determined in the previous section.

The general expression for transformation of the virtual spin $R$ operator around the $y$-axis is
\begin{eqnarray}
\mathbf{Y}_{02,13}(\varphi,f;\varphi _{1},g) = \mathbf{Y}_{02}(\varphi,f)\mathbf{Y}_{13}(\varphi _{1},g) = \cos(\varphi/2)(\mathbf{I}_{00}+\mathbf{I}_{22}) + \cos(\varphi _{1}/2)(\mathbf{I}_{00} +\mathbf{I}_{22}) \nonumber \\
+ \sin(\varphi/2)(\mathbf{I}_{20}e^{if}-\mathbf{I}_{02}e^{-if}) + \sin(\varphi _{1}/2)(\mathbf{I}_{31}e^{ig}-\mathbf{I}_{13}e^{-ig}). \label{equ3_30}
\end{eqnarray}
If $\varphi = \varphi _{1}$ and $f = g$ expression (\ref{equ3_30}) corresponds to equal rotation of virtual spin $R$ independently of spin $S$ states. The formulae (\ref{equ3_30}) transforms into the general expression $\mathbf{Y}_{02}(\varphi,f)\mathbf{Y}_{13}(\varphi _{1},g)$ for the rotation operator of virtual spin $S$ when the indices $1$ and $2$ of projective operators $\mathbf{I}_{mn}$ on the right hand side of (\ref{equ3_30}) are mutually replaced. The formula (\ref{equ3_30}) transforms in the general expression for rotation operator of virtual spin $R$ around the $x$-axis, if the phases are shifted: $f \rightarrow f - \pi/2, g \rightarrow g - \pi/2$. A simplified form of propagators without mentioning the above phases will be used below when the phases will be equal to zero.

Let us introduce a few ancillary operators, being constructed of pulse propagators
\begin{eqnarray}
\mathbf{L}_{mn,kl}(\alpha,\beta) = \mathbf{Y}_{mn,kl}(\pi/2,\pi/2)  \mathbf{X}_{mn,kl}(\alpha\pi/2,\beta\pi/2)\mathbf{Y}_{mn,kl}(-\pi/2,-\pi/2), \label{equ3_31}
\end{eqnarray}
\begin{displaymath}
\mathbf{M}_{mn}(\alpha) = \mathbf{Y}_{mn}(\alpha\pi) \mathbf{X}_{mn}(\pi),
\end{displaymath}
where  $\alpha,\beta = \pm 1$.

One can be convinced by means of simple multiplying of operators that the logic operations $\bm{\mathcal{P}}_{2} - \bm{\mathcal{P}}_{9}$ are expressed through the pulse propagator in the following manner
\begin{eqnarray}
\mathcal{P}_{2} & \equiv & \mathbf{NOT}_{1} = i \mathbf{X}_{02}(\pi)\mathbf{X}_{13}(\pi) = i \mathbf{X}_{02,13}(\pi,\pi), \nonumber \\
\mathcal{P}_{3} & \equiv & \mathbf{NOT}_{2} = i \mathbf{X}_{01}(\pi)\mathbf{X}_{23}(\pi) = i \mathbf{X}_{01,23}(\pi,\pi), \nonumber \\
\mathcal{P}_{4} & \equiv & \mathbf{NOT} = i \mathbf{X}_{03}(\pi)\mathbf{X}_{12}(\pi) = i \mathbf{X}_{03,12}(\pi,\pi), \nonumber \\
\mathcal{P}_{5} & \equiv & \mathbf{SWAP} = e^{-i\pi/2} \mathbf{L}_{01,23}(-1,1) \mathbf{X}_{12}(\pi), \nonumber \\
\mathcal{P}_{6} & \equiv & \mathbf{CNOT}_{1 \rightarrow 2} = e^{-i\pi/2} \mathbf{L}_{02,13}(-1,-1) \mathbf{X}_{23}(\pi), \nonumber \\
\mathcal{P}_{7} & \equiv & \mathbf{CNOT}_{2 \rightarrow 1} = e^{-i\pi/2} \mathbf{L}_{01,23}(-1,-1) \mathbf{X}_{13}(\pi), \nonumber \\
\mathcal{P}_{8} & \equiv & \mathbf{ICNOT}_{1 \rightarrow 2} = e^{-i\pi/2} \mathbf{L}_{02,13}(1,1) \mathbf{X}_{01}(\pi), \nonumber \\
\mathcal{P}_{9} & \equiv & \mathbf{ICNOT}_{2 \rightarrow 1} = e^{-i\pi/2} \mathbf{L}_{01,23}(1,1) \mathbf{X}_{02}(\pi) \label{equ3_32}
\end{eqnarray}
and similarly
\begin{eqnarray}
\mathbf{\Pi}_{0} & = & e^{-i\pi/2} \mathbf{L}_{02,13}(1,1) \mathbf{M}_{01}(1), \nonumber \\
\mathbf{\Pi}_{1} & = & e^{-i\pi/2} \mathbf{L}_{02,13}(1,1) \mathbf{M}_{01}(-1), \nonumber \\
\mathbf{\Pi}_{2} & = & e^{-i\pi/2} \mathbf{L}_{02,13}(-1,-1) \mathbf{M}_{23}(1), \nonumber \\
\mathbf{\Pi}_{3} & = & e^{-i\pi/2} \mathbf{L}_{02,13}(-1,-1) \mathbf{M}_{23}(-1). \label{equ3_33}
\end{eqnarray}
Existence of phase multipliers $i$ and $e^{-i \pi /2}$ does not lead to observable differences in quantum mechanics. Operators (\ref{equ3_32}) and (\ref{equ3_33}) are realized by using a pair of independent (not touching common energy levels) selective RF pulses.

Possible realization schemes for the pseudo Hadamard operator are
\begin{eqnarray}
_{1}\mathbf{h}_{R} = \mathbf{Y}_{01}(\pi/2)\mathbf{Y}_{23}(\pi/2), \nonumber \\
_{1}\mathbf{h}_{S} = \mathbf{Y}_{02}(\pi/2)\mathbf{Y}_{13}(\pi/2). \label{equ3_34}
\end{eqnarray}
It is constructed of $\pi/2$-pulses exciting independent transitions. The two qubit pseudo Hadamard operator is constructed of the operators (\ref{equ3_25}) according to the determination (\ref{equ2_19})
\begin{equation}
_{2}\mathbf{h} = ({_{1}\mathbf{h}}_{R})({_{1}\mathbf{h}}_{S}) = \mathbf{Y}_{01}(\pi/2)\mathbf{Y}_{23}(\pi/2)\mathbf{Y}_{02}(\pi/2)\mathbf{Y}_{13}(\pi/2). \label{equ3_35}
\end{equation}
It is easy to see that two qubit Hadamard operator can be realized by the pulse sequence
\begin{equation}
_{2}\mathbf{H} = {_{1}\mathbf{h}}_{R} \mathbf{Y}_{13}(2\pi) {_{1}\mathbf{h}}_{S} \mathbf{Y}_{23}(2\pi), \label{equ3_36}
\end{equation}
where  the operators $ _{1}\mathbf{h}$ and $ _{2}\mathbf{h}$ have to be used in the  form of  their expression (\ref{equ3_34}) through the pulse propogators.

\section{\label{sec:deutsch}Deutsch-Jozsa algorithm realization in  two pseudo spin information media}
The solution of the Deutsch-Jozsa  problem in a two qubit quantum system of a real spin $1/2$ pair was published by many authors \cite{deutsch1985,deutsch1992}. Here the solution of this problem will be given for the real spin $3/2$  Hilbert space which is presented as a direct product of two Hilbert spaces of virtual spins.

The Deutsch-Jozsa problem for a two qubit system consists in the following: let there exist a variable $x$ which takes two values $0$ and $1$, and let there be four functions $f_{mn}(x)$ depending on this argument $(m,n=0,1)$. The functions can be divided into two groups: \\
constant functions - $f_{00}(0)=f_{00}(1)=0,~ f_{11}(0)=f_{11}(1)=1$ \\
balanced functions - $f_{01}(0)=0, ~ f_{01}(1)=1, ~ f_{10}(0)=1, ~
f_{10}(1)=0$.

\begin{figure}
\includegraphics{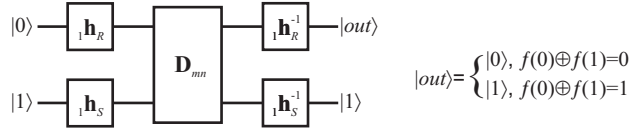}
\caption{\label{fig:f2} Quantum scheme of two-qubit Deutsch-Jozsa algorithm}
\end{figure}

One have to determine both meanings of the function at $x=0$ and $x=1$ to understand what class it belongs to. Quantum calculation allows to determine the function class by one measurement. In spite of the fact that this problem has no application meaning it plays an important role in quantum informatics since it had experimentally demonstrated the calculation acceleration due to the use of  the quantum laws. Implementation of the quantum scheme of the Deutsch-Jozsa algorithm is given in  Fig.~\ref{fig:f2}. The $\mathbf{D}_{mn}$ operators are in isomorphic correspondence to functions $f_{mn}(x)$. So the  identification of the operator $\mathbf{D}_{mn}$ properties is equivalent to determination the function $f_{mn}(x)$. One can observe that initial state $|01 \rangle $ is transformed into
\begin{equation}
\Psi^{*}= (-1)^{f(0)}[|0\rangle + (-1)^{f(0) \oplus f(1)} |1\rangle ][|0 \rangle - |1 \rangle ], \label{equ4_37}
\end{equation}
after the action of the logic operations ${_{1}\mathbf{h}}_{R},{_{1}\mathbf{h}}_{S}$ and $\mathbf{D}_{mn}$ shown in Fig.~\ref{fig:f2}. The following application of the Hadamard operator transforms  $|\Psi^{*} \rangle$  in  $|\Psi_{out} \rangle = | f_{mn}(0) \oplus f_{mn}(1),1 \rangle$ . Since the sum $f_{mn}(0) \oplus f_{mn}(1)$ equals to $0$ for the constant functions and equals to $1$ for the balanced functions, the measurement of the state of the first qubit in $|\Psi_{out} \rangle$ allows to determine the class of the function $f_{mn}(x)$.

\begin{figure}
\includegraphics{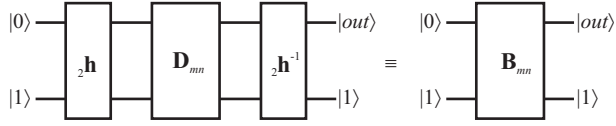}
\caption{\label{fig:f3}Equivalent quantum scheme of two-qubit Deutsch-Jozsa algorithm}
\end{figure}

It is useful to present the quantum scheme of the Deutsch-Jozsa algorithm (Fig.~\ref{fig:f2}) in the complete four-dimensional form (Fig.~\ref{fig:f3}).

One can obtain the operators $\mathbf{B}_{mn} = ({_{2}\mathbf{h}})\mathbf{D}_{mn}({_{2}\mathbf{h}^{-1}})$ following the last scheme and using the expressions (\ref{equ2_18}) and (\ref{equ2_17}):
\begin{eqnarray}
\mathbf{B}_{00} & = & \mathbf{P}_{00} + \mathbf{P}_{11} + \mathbf{P}_{22} + \mathbf{P}_{33} = \mathbf{E}, \nonumber \\
\mathbf{B}_{11} & = & \mathbf{P}_{00} + \mathbf{P}_{22} - \mathbf{P}_{11} - \mathbf{P}_{33}, \nonumber \\
\mathbf{B}_{01} & = & \mathbf{P}_{00} + \mathbf{P}_{22} + \mathbf{P}_{13} + \mathbf{P}_{31}, \nonumber \\
\mathbf{B}_{10} & = & \mathbf{P}_{00} + \mathbf{P}_{22} -
\mathbf{P}_{13} - \mathbf{P}_{31}. \label{equ4_38}
\end{eqnarray}

They have a sense of the Deutsch-Jozsa problem solution in a form of a single operator. The direct application of  the $\mathbf{B}_{mn}$ operators to the initial state $|01 \rangle$ gives the result
\begin{eqnarray}
\mathbf{B}_{00}|0,1\rangle = |0,1\rangle, & \mathbf{B}_{11} |0,1\rangle = -|0,1\rangle, \nonumber \\
\mathbf{B}_{01}|0,1\rangle = |1,1\rangle, & \mathbf{B}_{10} |0,1\rangle = -|1,1\rangle. \label{equ4_39}
\end{eqnarray}

So the operators $\mathbf{B}_{00}$ and $\mathbf{B}_{11}$ corresponding to the constant functions does not change the initial state $|0,1 \rangle$ and operators $\mathbf{B}_{01}$ and $\mathbf{B}_{10}$ corresponding to the balanced ones transform it into $|1,1 \rangle$, as it was determined before.

Here an accelerated implementation of the Deutsch-Jozsa algorithm can be proposed. The acceleration idea consists of the experimental implementation of the final operations $\mathbf{B}_{mn}$ (\ref{equ4_38}) instead of consecutive realization of all processors of the standard scheme (Fig.~\ref{fig:f2}). This realization can be reached in the following way
\begin{eqnarray}
\mathbf{B}_{00} & = & \mathbf{E} = \mathbf{P}_{00} + \mathbf{P}_{11} + \mathbf{P}_{22} + \mathbf{P}_{33}, \nonumber \\
\mathbf{B}_{11} & = & \mathbf{X}_{13}(2\pi) = \mathbf{P}_{00}+ \mathbf{P}_{22} - (\mathbf{P}_{11}+\mathbf{P}_{33}), \nonumber \\
\mathbf{B}_{01} & = & \mathbf{X}_{13}(\pi)= \mathbf{P}_{00} + \mathbf{P}_{22} - i(\mathbf{P}_{13} + \mathbf{P}_{31}), \nonumber \\
\mathbf{B}_{10} & = & \mathbf{X}_{13}(-\pi) = \mathbf{P}_{00} + \mathbf{P}_{22} + i(\mathbf{P}_{13} + \mathbf{P}_{31}). \label{equ4_40}
\end{eqnarray}

Comparing (\ref{equ4_40}) and (\ref{equ4_38}) one can see that the realization scheme does not fully coincides with the requirements of the mathematical logic because of the imaginary unit in off-diagonal matrix elements. To estimate this difference let us consider how operations (\ref{equ4_40}) influence the states of  the two qubit system
\begin{eqnarray}
\mathbf{B}_{00}|0,1\rangle = |0,1\rangle, & \mathbf{B}_{11} |0,1\rangle = -|0,1\rangle, \nonumber \\
\mathbf{B}_{01}|0,1\rangle = -i|1,1\rangle, & \mathbf{B}_{10} |0,1\rangle = i|1,1\rangle. \label{equ4_41}
\end{eqnarray}

So in comparison with standard demands (\ref{equ4_39}) in quantum realization (\ref{equ4_41}) of the Deutsch-Jozsa algorithm the final state $|\Psi_{out} \rangle$ turns out to be multiplied by the phase factor $e^{-i\pi/2}$ in the case of the  balanced function $f_{mn}(x)$. As was mentioned above this fact does not influence observation results in quantum mechanics.

How can we prepare the initial state $|0,1 \rangle$  required for the beginning of the calculation if we start from a thermodynamically equilibrium density matrix $\bm{\rho}_{T}$, which corresponds to a mixed state of spin $3/2$? Probably the most simply it can be done by applying the following pulse sequence
\begin{displaymath}
\bm{\mathcal{P}} = \mathbf{X}_{02}(\pi/2)\mathbf{X}_{23}(\pi)\mathbf{G},
\end{displaymath}
where $\mathbf{G}$ is the pulsed magnetic field gradient
\cite{kchitrin2000}. $\mathbf{G}$-pulse damps all off-diagonal
matrix elements of the density matrix and transforms the density
matrix $\bm{\rho}_{T}$ into that of  $\bm{\rho}_{qp} = \mathbf{E}
+ \alpha \mathbf{I}_{11}$ for the pseudo-pure state, where
$\alpha$ is a parameter depending on the temperature and Larmour
frequency. The unity operator $\mathbf{E}$ can be omitted in
$\bm{\rho}_{pq}$ since it is not changed under any unitary
transformation produced by pulse sequences and since the mean
value of it's product $\mathbf{E} \times \mathbf{I}_{\lambda}$
with any spin operator $\mathbf{I}_{\lambda}$ equals to zero. So
the density matrix $\bm{\rho}_{pq}$ turns out to be equivalent to
the required initial density matrix $\bm{\rho}_{initial} = \alpha
\mathbf{I}_{11}$ of the pure state $|0,1 \rangle $.

For the read out the calculation result it is enough to apply the
selective $\pi/2$-pulse to the transition $\hbar \varepsilon _{1}
\leftrightarrow  \hbar \varepsilon _{2}$. The existence of a free
induction signal produced by this pulse will demonstrate that
after the calculation the spin system occurs in the state
$|\Psi_{out} \rangle = |1,1 \rangle $, that is we have determined
the balanced function. The absence of the free induction signal
will demonstrate that  $|\Psi_{out} \rangle = |0,1 \rangle $ and
that we have a deal with the constant function.

\section{Final remarks}

1. All above obtained results are applicable for every four
arbitrarily chosen energy levels of any physical system. It is
only important to have a possibility to excite (as simply as
possible) resonance transitions required for the quantum
processors realization.  NMR systems were chosen here for
consideration in detail because it is namely in NMR pulse
sequences have for a long time been excellently theoretically
derived and used for investigation of spin dynamics and kinetics.
A discrete optical state of atoms in solids as an example of two
qubit medium of other physical nature were considered in
\cite{kess1999b}.

2. The experimental realization of the logical processors schemes in NMR proposed in Sec.~\ref{sec:experimental} are not unique, they are rather the simplest. Without anisotropy in the plane, perpendicular to the dc magnetic field, the $x$ and $y$ directions are equivalent and this fact gives additional possibility for processor realization by means of RF fields being oriented along any of these axis.

3. Another reason of an implementation schemes variety arises from a possibility to excite   transitions with different selection rules: $\delta m = |\Delta m| = 1,2, \ldots $ Though all resonance transition in the spin spectrum of quadrupole nuclei placed in low symmetric crystalline field are allowed in principle, their probabilities decrease as $(\omega _{q}/ \omega _{0})^{2 \delta m-2}$ or $(\omega _{q} \eta /\omega _{0})^{2 \delta m-2}$ with growth of $ \delta m$ where $\omega _{0}$ and $ \omega _{q}$ are Larmour frequency and its quadrupole shift, $\eta$-crystalline field asymmetry parameter $(0 \leq \eta \leq 1)$. For this reason we proposed (where it is possible) to excite $ \delta m =1 $ transitions. Never the less some implementation schemes in sections \ref{sec:experimental} and \ref{sec:deutsch} contain  $m = 2$ and  $m = 3$ transitions. One suitable method of exiting  $m = 2$ transitions was utilized in paper \cite{kchitrin2000} for creation the pseudo pure states of quadrupole spin $3/2$. However there is a more direct way, unfortunately being connected with rapid grow of the sequences length:
\begin{eqnarray}
X_{03}(\varphi) = X_{01}(-\pi)X_{12}(-\pi)X_{23}(-\varphi)X_{12}(\pi)X_{01}(\pi) = Y_{02}(-\pi)X_{23}(\varphi)Y_{02}(\pi)
\nonumber \\
= Y_{01}(-\pi)Y_{12}(-\pi)X_{23}(\varphi)Y_{12}(\pi)Y_{01}(\pi), \nonumber \\
Y_{03}(\varphi) = X_{01}(-\pi)X_{12}(-\pi)Y_{23}(-\varphi)X_{12}(\pi)X_{01}(\pi) = Y_{02}(-\pi)Y_{23}(\varphi)Y_{02}(\pi) \nonumber \\
= Y_{01}(-\pi)Y_{12}(-\pi)Y_{23}(\varphi)Y_{12}(\pi)Y_{01}(\pi).
\end{eqnarray}

4. For going over to the virtual spin representation it is not necessary that direct products of the virtual spin states $|00 \rangle, |01 \rangle, |10 \rangle$ and $|11 \rangle$ were arranged in the consecutive order in an energetic scale, as in Fig.~\ref{fig:f1}. The arrangement order can arbitrarily be dictated by convenience of the gates implementation. In particular there may be intermediate physical states not participating in information processes \cite{kess1999b}.

5. Off-diagonal matrix elements of the main quantum gates (\ref{equ2_7}) - (\ref{equ2_16}) have to be positive real quantities (unities) according to the algorithm theory. Their experimental implementation in NMR meets a problem: evolution operators of the real RF pulse sequences contain either imaginary (\ref{equ3_25}) or negative (\ref{equ3_26}) off-diagonal matrix elements.

In some cases $(\bm{\mathcal{P}}_{2}, \bm{\mathcal{P}}_{3}, \bm{\mathcal{P}}_{4})$ a pulse sequence scheme can be arranged in such a way that the gates are realized rather simple except for the phase factor $e^{i \alpha}$, which as was mentioned above, does not influence  the calculation results. In other cases more complicated sequences are necessary. This property is not an attribute of qubit encoding in the virtual spin representation and is inherent to every method of qubit encoding. Some rare cases are known in which the imaginary off-diagonal matrix element existence does not change  the calculation results (\cite{jones2000}, chapter 4).

It may be useful to adapt the following special receptions for simplification of realization schemes of algorithms:

A. The use the processors which properties does not exactly coincide in the form with demands of algorithms theory. A typical example - the pseudo Hadamard operator $_{2}\mathbf{h}$ was for a long time used in quantum informatics instead of Hadamar operator $_{2}\mathbf{H}$. Going this way, we can propose a simplified realization of some quantum processors
\begin{eqnarray}
\bm{\mathcal{P}}_{5} \Rightarrow {^{*}\bm{\mathcal{P}}}_{5}  =  X_{12}(\pi), & \bm{\mathcal{P}}_{6} \Rightarrow {^{*}\bm{\mathcal{P}}}_{6}  =  X_{23}(\pi), \nonumber \\
\bm{\mathcal{P}}_{7} \Rightarrow {^{*}\bm{\mathcal{P}}}_{7}  =  X_{13}(\pi), & \bm{\mathcal{P}}_{8} \Rightarrow {^{*}\bm{\mathcal{P}}}_{8}  =  X_{01}(\pi), \nonumber \\
\bm{\mathcal{P}}_{9} \Rightarrow {^{*}\bm{\mathcal{P}}}_{9}  =  X_{02}(\pi), \label{equ5_42}
\end{eqnarray}
where an operator~ $^{*}\bm{\mathcal{P}}_{m}$ is the operator $\bm{\mathcal{P}}_{m}$, which off-diagonal matrix elements are multiplied by the imaginary unit $i$. Processor realization in the form of~ $^{*}\bm{\mathcal{P}}_{m}$ was used in \cite{valiev2001} for simplicity and for concentration of  attention  on the virtual spin representation.

B. The final state $|\Psi _{out} \rangle$  resulting after the simplified gates application can not coincide with demands of quantum calculations theory. It is quite enough to initially establish the isomorphic correspondence between $|\Psi _{out} \rangle$ and desired object properties (function class in the Deutsch-Jozsa algorithm, particularities of a desired object from a non-ordered collection as in the Grover's algorithm). As an example of such an approach the Deutsch-Jozsa algorithm realization scheme derived in the previous section can be served.

C. Realization of algorithms in the form of a single processor (Fig.~\ref{fig:f3}) instead of the consecutive realization of each of the quantum processors presented in Fig.~\ref{fig:f2} may frequently turn out to be more simple. The Grover's algorithm realization scheme in virtual spin representation was derived in this manner and will be send in print in the nearest future.


\end{document}